\documentclass{procDDs}
\usepackage{amssymb}
\usepackage{color,graphicx}
\usepackage{amsmath} %

\def\bc{\begin{center}}
\def\ec{\end{center}}
\def\beq{\begin{equation}}
\def\eeq{\end{equation}}

\def\tg{{\textrm{tg}\,}}
\def\pial{{\frac{\pi\alpha(t)}{2}}}
\def\pialz{{\frac{\pi\alpha(0)}{2}}}
\def\pit{{\frac{\pi}{2}}}
\def\ii{{{\rm i}}}

\def\RE{{{\rm Re}}}

\def\tg{{{\rm tg}}}
\def\alp{{\alpha'}}

\def\kperp{{\textbf{k}_{\perp}}}
\def\kp#1{{\textbf{k}_{#1\perp}}}
\def\kps#1{{\textbf{k}^2_{#1\perp}}}

\title{Restrictions on pp scattering amplitude\\
imposed by first diffraction minimum data\\ obtained by TOTEM at LHC}

\author{\textbf{V.N. Kovalenko}, A.M. Puchkov, V.V. Vechernin, D.V. Diatchenko}% List of authors with the
{Dep. of High Energy Physics, Saint-Petersburg State University, Russia}                 % same affiliation,
{nvkinf@rambler.ru}                                   % e-mail of lecturer
\begin {document}

\maketitle

\index{Kovalenko, V.N.}                              % write this for each author
\index{Puchkov, A.M.}                              % to generate the index
\index{Vechernin, V.V.}                             %
\index{Diatchenko, D.V.}                             %
%generate the index

\begin{abstract}%
We present the systematic analysis of the behavior of elastic pp scattering
cross section in the region of the diffraction peak and the first diffraction
minimum in the framework of quasi-eikonal Gribov-Regge approach,
taking into account the unitarity condition.
We use the formalism taking into consideration the dependence of pomeron
signature factor on $t$. We show that although in this approach
one can describe the general features of pp scattering,
the behavior of pp elastic differential cross section
in the vicinity of the first diffraction minimum
measured by the TOTEM experiment at LHC can not be reproduced
under any parameter values at standard assumptions of the model.
Physically, in the quasi-eikonal Regge approach the proton
proved to be ``gray'' and rather large, whereas the TOTEM
data indicate that it is smaller
and closer to ``black'' at LHC energies.

%We present the systematic analysis of the behavior of elastic pp scattering cross section
%in the region of the diffraction peak and the first diffraction minimum in the framework of
% quasi-eikonal Regge approach, taking into account the unitarity condition. We extend the formalism in order to take into account the dependence of pomeron signature factor on $t$.
%We show that although in this approach
%one can describe the general features of the pp scattering, the behavior of pp elastic differential cross section in the vicinity
%of the first diffraction minimum measured by the TOTEM
%experiment at LHC can not be reproduced
%under any parameter values at standard assumptions of the model.
%Physically one can say that  in quasi-eikonal Regge approach the nucleon
%proved to be ``gray'' and rather large, whereas the TOTEM
%experiment indicatives that it is smaller and closer to ``black'' one.
%%This conclusion is consistent with the results, obtained by the 
%%modeling of impact parameter profile function in \cite{Dremin1}
%%using the TOTEM data.
\end{abstract}
\section{Introduction}%
Recently, fine experimental data on the behavior of elastic scattering cross section
in the region of the diffraction peak and the first diffraction minimum was obtained by the TOTEM
experiment for 7 TeV pp collisions at the LHC \cite{TOTEM11, TOTEM13}.
The first restrictions on pp scattering amplitude imposed by this data were discussed in \cite{Dremin0,Dremin1,Dremin2,Dremin3,Uzhinsk12}.
In the present work we consider the problem in the framework of
%dipole model [5]  and
quasi-eikonal Gribov-Regge approach \cite{Gribov67, TerMart72} taking into account the unitarity condition.
We show that although in this approach
one can describe the general features of the pp scattering
such as the total elastic and inelastic cross sections, the slope of diffraction cone and the multiplicity density
at mid-rapidities \cite{CapFer131,CapFer132},
nevertheless the behavior of pp elastic differential cross section in the vicinity
of the first diffraction minimum measured by the TOTEM
experiment at LHC \cite{TOTEM11, TOTEM13} can not be reproduced
under any parameter values at standard assumptions of the model.
%Physically one can say that  in quasi-eikonal Regge approach the nucleon
%proved to be ``gray'' and rather large, whereas the TOTEM
%experiment indicatives that it is smaller and closer to ``black'' one.
%This conclusion is consistent with the results, obtained by the
%modeling of impact parameter profile function in \cite{Dremin1}
%using the TOTEM data.

\section{Quasi-eikonal Gribov-Regge approach}
The Gribov-Regge approach \cite{Gribov67} is aimed  to describe the high-energy hadronic interactions
in the soft domain,  when $s=(p_1+p_2)^2\to \infty$ and  $t=(p_1-p'_1)^2 \lesssim m^2$.
Here $p_1$, $p_2$ and $p'_1$ are the four-momenta of colliding protons and the scattered hadron.
For pp elastic interactions in the center mass system (cms): $\sqrt{s}=2 E$ and  $t=-k^2$,
where $2E$ and $k$ are the total energy and the transferred momentum in cms
($k\approx|\textbf{k}_{\perp}|$ in this limit).
In this approach in the amplitude of elastic pp scattering at high energy
the exchange by the reggeon with vacuum quantum numbers  - the pomeron - is dominated.
The amplitude corresponding to the one pomeron exchange is given by the following expression \cite{Gribov67}:
\begin{equation} \label{A1}
A_1(s,t)= G(t)\ D(s,t) \ ,
\end{equation}
where $D(s,t)$ is the pomeron propagator:
\begin{equation} \label{D}
D(s,t)=\eta(\alpha(t))\  \left(\frac{s}{s_0}\right)^{\alpha(t)-1} \ .
\end{equation}
Here $\alpha(t)$ is the pomeron trajectory:
\begin{equation} \label{al}
\alpha(t)=1+\Delta+\alpha' t=1+\Delta-\alp k^2,
\end{equation}
and $\eta(\alpha(t))$ is the so-called signature factor:
\begin{gather}
\label{eta}
\eta(\alpha(t))=-\frac{\exp(-\ii \pial)}{\sin(\pial)}
%=\frac{-\cos(\pial)+\ii \sin(\pial)}{\sin(\pial)}=\ii-\ctg(\pial)=
=\\
\nonumber
=\ii+\tg\left(\pit (\alpha(t)-1)\right)=\ii+\tg\left(\pit (\Delta-\alp k^2)\right)        \ .
\end{gather}
The $s_0$ is a parameter of the order of hadronic masses. Usually, one puts $s_0=1 \mathrm{\ GeV}^2$ and
writes the pomeron propagator (\ref{D}),
introducing the whole rapidity width: $y\equiv \ln(s/s_0)$,
in the following form:
\begin{equation} \label{pom}
D(s,t)=\eta(\alpha(t))\  \exp(\Delta y)\exp(-\alp y k^2) \ .
\end{equation}
The factor $G(t)$ in (\ref{A1}) originates from the cut vertexes
of pomeron coupling to protons. Usually, one approximates it
by a Gaussian:
\begin{equation} \label{G}
G(t)=G_0\exp(-R^2 k^2) \ .
\end{equation}

The Froissart bound at $s\to\infty$, applied to (\ref{A1}), leads to the restriction: $\alpha(0)\leq 1$,
but the experimental data on the increase of the total cross sections and the multiplicity density
at mid-rapidities with energy demands $\alpha(0)> 1$,
so one has to consider the supercritical pomeron with intercept $\alpha(0)-1=\Delta> 0$.
To meet the Froissart bound in this case one has to take into account
the exchange by an arbitrary number of pomerons in the elastic pp amplitude:
\begin{equation} \label{A}
A(s,t)=\sum_{n=1}^{\infty} A_n(s,t) \ ,
\end{equation}
where $A_n(s,t)$ is the amplitude corresponding to the exchange by $n$ pomerons \cite{Gribov67}:
\begin{gather}  \nonumber
A_n(s,t)= \\ \nonumber
=\frac{\ii^{n-1}}{\pi^{n-1}n!}\int \prod_{i=1}^{n} d\kp i D(s,\kps i)\  G_n(\kp 1,...,\kp n) \times \ \\
\times \delta^{(2)}(\sum_{i=1}^n \kp i - \kperp) \ . \label{An}
\end{gather}
Usually, one supposes the factorization of the cut vertex
of $n$ pomerons coupling to proton:
\begin{equation} \label{Gn}
G_n(\kp 1,...,\kp n)=C_n  \prod_{i=1}^{n} G(\kps i) \ .
\end{equation}
Assuming additionally
\begin{equation} \label{Cn}
C_n=C^{n-1},
\end{equation}
one comes to the quasi-eikonal Regge approximation for the pp amplitude.

\section{Taking into account the dependence of pomeron signature factor on $t$}%

Frequently, one replaces the signature factor (\ref{eta}) in the pomeron propagator (\ref{pom})
simply by $\ii$  (see e.g. \cite{Werner}).
In this appro\-xi\-ma\-tion the amplitude of one pomeron exchange  is pure imaginary.
The present paper is aimed to investigate the behavior of the differential elastic pp cross section
in the vicinity of first diffraction minimum at $-t=k^2_{min}$, so
 we need to take into account the dependence of this factor on $t$.
 For this purpose we'll use the approximation suggested in \cite{Volkovitskiy}:
\begin{gather}
\nonumber
\eta(\alpha(t))\approx-\frac{\exp(-\ii \pial)}{\sin(\pialz)}=
\eta(\alpha(0))\exp(\ii \pit \alp k^2)= \\
=
\left[\ii+\tg(\pit \Delta)\right]\exp(\ii \pit \alp k^2)
%=\frac{-\cos(\pial)+\ii \sin(\pial)}{\sin(\pial)}=\ii-\ctg(\pial)=
  \ , \label{eta1}
\end{gather}
which is valid when $|\Delta|,|\alp k^2|\ll 1$ and hence $|\alpha(t)-1|=|\Delta-\alp k^2|\ll 1$.
The advantages of this approximation are that it has a Gaussian form and in the first order it coincides
with the expansion of exact signature factor (\ref{eta}):
\beq
\label{equiv}
\eta(\alpha(t))\approx\ii+\pit\Delta-\pit\alp k^2         \ .
\eeq
Note that in accordance with (\ref{D}) and (\ref{eta}) the corrections affect only the real part
of pomeron propagator and do not change its imaginary part.

The factorization suggestion, (\ref{Gn}), and the Gaussian form of (\ref{G}) and (\ref{eta1})
enable to calculate the amplitudes $A_n(s,t)$, (\ref{An}), explicitly:
\begin{gather}\nonumber
A_n(s,t)= \ii \, C_n \left[1-\ii\,\tg(\pit \Delta)\right]^n \frac{G_0 \exp (\Delta y)}{n\,n!}
\times \\
\times \left( \frac{ -G_0 \exp (\Delta y)}{R^2+\alp y'} \right)^{n-1}
 \exp\left[-\frac{R^2+\alp y'}{n}k^2\right],
  \label{An1}
\end{gather}
where $y'=y-\ii\pi/2$.  Assuming additionally the quasi-eikonal approximation for  $C_n$, (\ref{Cn}),  one has
\begin{gather} \nonumber
A_n(s,t) = \ii \,  \left[1-\ii\,\tg(\pit \Delta)\right]^n \frac{G_0 \exp (\Delta y)}{n\,n!}
\times \\
\times
\left( \frac{ -C\, G_0 \exp (\Delta y)}{R^2+\alp y'} \right)^{n-1}
 \exp\left[-\frac{R^2+\alp y'}{n}k^2\right] \ . \label{An2}
\end{gather}
Now summing over $n$, (\ref{A}), one can calculate the amplitude $A(s,t)$
and hence the elastic differential pp cross section:
\begin{equation} \label{dsdt}
\frac{d\sigma_{el}}{dt}=4\pi|A(s,t)|^2 \ .
\end{equation}

\section{Comparison with the experimental data}
\begin{figure}[t!]\centering
  \includegraphics[width=.49\textwidth]{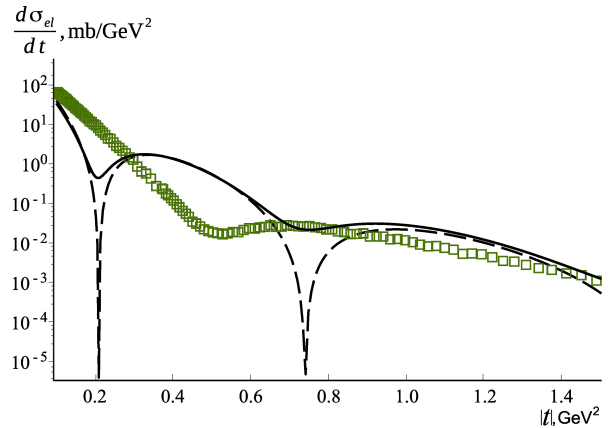}
\caption{ Elastic differential cross section of pp scattering,
calculated
for parameters ${C=1.8}$ and ${R^2=3.3}$~GeV$^{-2}$,
in two versions of the model:
with, (\ref{eta1}), (solid line) and without, $\eta=\ii$, (dashed line)
taking into account a dependence of the signature factor $\eta(\alpha(t))$, (\ref{eta}), on $t$.
 The points are the TOTEM experimental data \cite{TOTEM11,TOTEM13}.}
\label{fig1}
\end{figure}

In paper \cite{CapFer132} it is shown that
the general features of the pp scattering at the LHC energy
such as the total cross sections, the slope of diffraction cone and the multiplicity density
at mid-rapidities can be reproduced in the quasi-eikonal Regge approach
under the following values of parameters:
%\begin{equation}
\begin{gather}
\label{FCparam}
\nonumber
\Delta= 0.19,\ \alpha'=0.25\ \mathrm{GeV}^{-2},\ G_0=0.85\ \mathrm{GeV}^{-2}, \\
\ \  R^2=3.3\ \mathrm{GeV}^{-2},\ C=1.8,\ \mu_0=1.5 \  ,
\end{gather}
where $\mu_0$ is the multiplicity at mid-rapidities from one cut pomeron.
So with these values of parameters we calculate by (\ref{A}) and (\ref{An2})
the elastic differential pp cross section, (\ref{dsdt}),
as a function of cms transfer momentum
and compare the obtained results with the corresponding experimental data,
measured by the TOTEM experiment for 7 TeV pp collisions at the LHC \cite{TOTEM11,TOTEM13}.
The calculation were fulfilled with the signature factor both given by (\ref{eta1}) and
$\eta(\alpha(t))=\ii$. The comparison is presented in Fig.~\ref{fig1}.
We see that although this set of parameters satisfactorily describes
the total cross sections, the slope of diffraction cone, $B$, and the multiplicity density
at mid-rapidities (see Table~\ref{TableN1}), nevertheless it can not reproduce the behavior of
elastic differential pp cross section in a vicinity
of the first diffraction minimum measured by the TOTEM
experiment.

\begin{table}[b]
%\begin{tabular}{|y{1.2cm}|y{3.7cm}|y{7.4cm}|y{3.4cm}|}%{|c|c|c|c|}
\begin{tabular}{|c|c|c|c|}
%\begin{tabular}{| >{\centering\arraybackslash}m{1.40cm} | >{\centering\arraybackslash}m{1.35cm}| >{\centering\arraybackslash}m{1.45cm}| >{\centering\arraybackslash}m{1.75cm}|}
\hline
& $\eta$, (\ref{eta1})& $\eta=\ii$ &TOTEM data \\ \hline
$\sigma_{inel}$, mb & 78.73 & 78.54 & 73.15$\pm$1.26 \\ \hline
$\sigma_{tot}$, mb & 98.86 & 97.56 & 98.58$\pm$2.23 \\ \hline
$\sigma_{el}$, mb & 20.12 & 19.02 & 25.43$\pm$1.073 \\ \hline
$B$, $\mathrm{GeV}^{-2}$ & 23.53 & 23.41 & 19.9$\pm$0.3 \\ \hline
\end{tabular}
\caption{General characteristics (see the text) of pp scattering, calculated
for parameters $C=1.8$ and ${R^2=3.3}$~GeV$^{-2}$,
in two versions of the model:
with, (\ref{eta1}), and without, $\eta=\ii$, taking into account a dependence of the signature factor
$\eta(\alpha(t))$, (\ref{eta}), on $t$,
compared with the experimental data \cite{TOTEM11,TOTEM13}.}
\label{TableN1}
\end{table}

\begin{table}[b!]
\begin{tabular}{|c|c|c|c|}
%\begin{tabular}{|y{1.2cm}|y{3.7cm}|y{7.4cm}|y{3.4cm}|}%{|c|c|c|c|}
%\begin{tabular}{| >{\centering\arraybackslash}m{1.40cm} | >{\centering\arraybackslash}m{1.35cm}| >{\centering\arraybackslash}m{1.45cm}| >{\centering\arraybackslash}m{1.75cm}|}
\hline
% & Model with sig. factor & Model, constant sig. factor & Experiment \\ \hline
& $\eta$, (\ref{eta1})& $\eta=\ii$ &TOTEM data \\ \hline
$\sigma_{inel}$, mb & 66.32 & 67.50 & 73.15$\pm$1.26 \\ \hline
$\sigma_{tot}$, mb & 104.07 & 102.99 & 98.58$\pm$2.23 \\ \hline
$\sigma_{el}$, mb & 37.75 & 35.49 & 25.43$\pm$1.073 \\ \hline
$B$, $\mathrm{GeV}^{-2}$ & 14.06 & 14.01 & 19.9$\pm$0.3 \\ \hline
\end{tabular}
\caption{The same as in Table \ref{TableN1}, but calculated for parameters $C=1.0$ and ${R^2=0.3}$~GeV$^{-2}$.}
\label{TableN2}
\end{table}

Note that the position of the minimum, $|t_{dip}|=k^2_{min}\approx0.5\,\mathrm{GeV}^{2}$,
is in the domain where using of the approximation (\ref{eta1}) are valid,
since ${\Delta=0.19\ll 1}$ and ${\alpha'k^2_{min}\approx 0.13\ll 1}$.

\begin{figure}[t!]\centering
  \includegraphics[width=.49\textwidth]{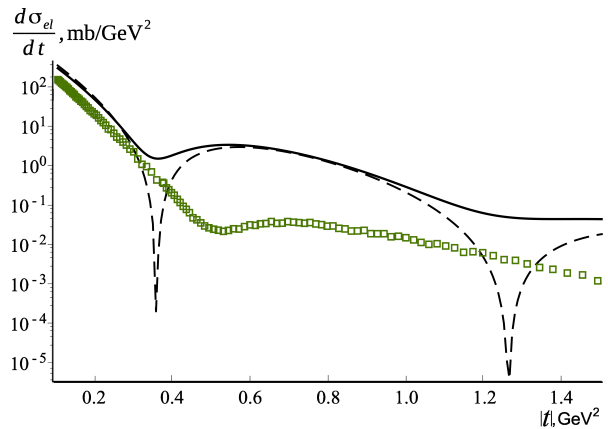}
  \vspace{-0.63cm}
\caption{The same as in Fig.~\ref{fig1}, but calculated for parameters $C=1.0$ and ${R^2=0.3}$~GeV$^{-2}$.}
\label{fig2}
\end{figure}

\begin{figure}[t!]\centering
  \includegraphics[width=.42\textwidth]{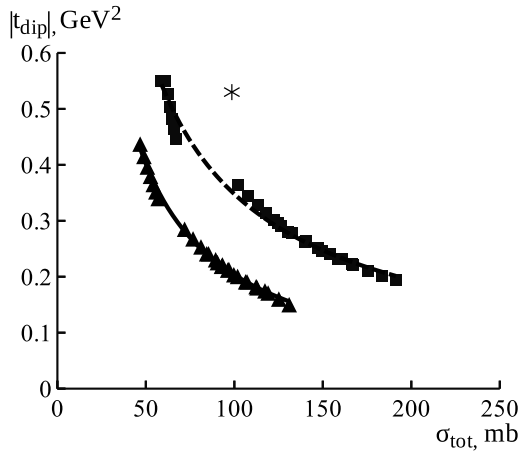}
\caption{The position of the first diffraction minimum, $|t_\mathrm{dip}|$, and the value of total cross section,
$\sigma_{\mathrm{tot}}$, calculated for parameters C=1.8 (triangles) and C=1 (squares),
with different values of $R^2$ and $G_0$ .
The lines are approximations by the power law. % ($|t_{dip}|=const/{\sigma_{tot}}^\alpha$).
The TOTEM experimental data is marked by an asterisk.
} \label{fig3}
\end{figure}

Moreover, the following analysis shows that the TOTEM data
can not be reproduced under \textbf{{any}} parameter values in the framework
of the quasi-eikonal Regge approach under standard assumptions.
%Really,
Actually,
by (\ref{An2}) we see that at given energy and fixed quasi-eikonal parameter $C$
we have only two effective parameters: $R^2+\alpha' y$ and $G_0\exp(\Delta y)$.
So we have calculated the $\sigma_{tot}$ and the position, $|t_{dip}|=k^2_{min}$,
of the first diffraction minimum in $d\sigma_{el}/dt$
varying independently the parameters $R^2$ and $G_0$. The other variables were fixed
as follows: $\Delta= 0.19$, $\alpha'=0.25\ \mathrm{GeV}^{-2}$, $C=1.8$.
The results are presented in Fig.~\ref{fig3}. We see that one can not describe simultaneously the total cross section
and the position of the diffraction minimum in the framework of the model.
All variation results are strongly correlated and lay along one curve,
approximately described by the power law: $|t_{dip}|=const/\sigma_{tot}$.

The situation is better in the simple eikonal approximation with $C=1$,
but even in this case one can not obtain  simultaneously the values $\sigma_{tot}$
and $|t_{dip}|=k^2_{min}$  as in the TOTEM experiment,
see Fig.~\ref{fig3}.
The general characteristics of pp scattering for this case are presented in Table~\ref{TableN2} and Fig.~\ref{fig2}.

\begin{figure}[t!]\centering
  \includegraphics[width=.3169\textwidth]{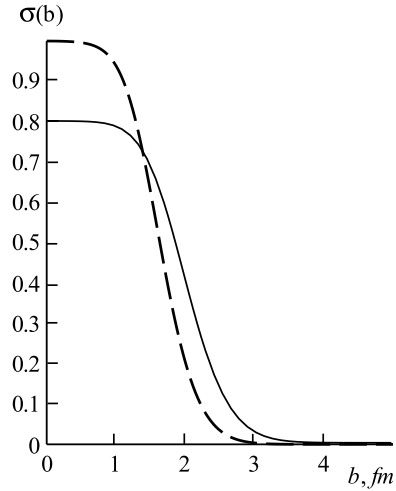}
\caption{Profile function of pp collision $\sigma(b)$ calculated at $C=1.8$ (solid line) and ${C=1}$ (dashed line).
$\sigma(b){=2\RE( \gamma(b)) - |\gamma(b)|^2}$, where $\gamma(b)$, (\ref{gamma}),
is a Fourier transform of the amplitude.} \label{fig4}
\end{figure}

Physically one can understand this behaviour recalling that in the
quasi-eikonal approach the impact parameter profile function $\gamma(s,b)$
of pp interaction (the Fourier transform of the $-2\,\ii A(s,t)$)
 has the form:
\beq
\label{gamma}
\gamma(s,b)=C^{-1}[1-\exp(C\omega(s,b))] \ ,
\eeq
where $\omega(s,b))$ is the Fourier transform of the one pomeron exchange amplitude $-2\,\ii A_1(s,t)$, (\ref{A1}).
At large energy in the model with supercritical pomeron at $b=0$ we have
$\RE\, \omega(s,0)\gg 1$ and hence $\gamma(s,0) \approx 1/C$.
Taking this into account
we see in Fig.~\ref{fig4}  that  in quasi-eikonal Regge approach at $C=1.8$ the nucleon
proved to be ``gray'', $1/C<1$, and rather large in the impact parameter space, which leads to smaller value
of the first diffraction minimum  position ($\approx0.2\,\mathrm{GeV}^{2}$) in the momentum space, see Fig.~\ref{fig1}.
Whereas the TOTEM experiment by larger value of the first diffraction minimum position,
$|t_{dip}|=k^2_{min}\approx0.5\,\mathrm{GeV}^{2}$,
indicates that the nucleon is smaller in the impact parameter space
and closer to ``black'', what corresponds
to better description with $C=1$ than with $C=1.8$, in Fig.~\ref{fig3}.

The dependencies in Fig.~\ref{fig3} correspond to the fact that at given
transparency, $C$, the increase of nucleon radius corresponds
to the rise of total cross section and to the decrease of the Fourier transform width in momentum space.
So by the comparison with the TOTEM data in Fig.~\ref{fig1} we see that
since the data has the larger value of the first diffraction minimum position,
than in quasi-eikonal Regge approach, then it corresponds
to the smaller interaction radius and hence to more ``black'' nucleon
to ensure the same total pp cross section.

This conclusion is in correspondence with the results, obtained by the
modeling of the nucleon impact parameter profile function in \cite{Dremin1}
using the TOTEM data.

% and results of our modeling of the behavior of the differential elastic scattering cross section
%as a function of transfer momentum, using the impact parameter profile function as a starting point.
%The results are shown in Fig.4. They confirm the above qualitive consideration.

\section{Conclusions}%
We come to the conclusion that
in the framework of the
quasi-eikonal Regge approach \cite{Gribov67, TerMart72}
under standard assumptions
one can not describe the behavior of pp elastic differential cross section in the vicinity
of the first diffraction minimum measured by the TOTEM experiment at LHC \cite{TOTEM11, TOTEM13}
under any parameter values,
although one can describe the general features of the pp scattering
such as the total elastic and inelastic cross sections, the slope of diffraction cone and the multiplicity density
at mid-rapidities \cite{CapFer131,CapFer132}.

So to adjust the Regge model to the new experimental data we have to reconsider
some standard simplifying assumptions of the model,
namely,
%. They are
the factorization of the cut vertex of $n$ pomerons coupling to proton, (\ref{Gn}),
the quasi-eikonal assumption, (\ref{Cn}), and
the Gaussian form of the cut vertex of one pomeron coupling to proton, (\ref{G}).

Most simply one can give up the latter assumption.
%The most simple solution is just to give up the latter assumption.
This is also supported by the latest experimental data of the TOTEM collaboration \cite{TOTEM15}.
In the paper the authors argue that the first diffraction peak in pp scattering at 8 TeV
has the deviation from the simple Gaussian form, {$\exp(Bt)=\exp(-Bk^2)$},
and is better described by the dependency $\exp(b_1t+b_2t^2)$ or even by $\exp(b_1t+b_2t^2+b_3t^3)$.
Clearly that we can reproduce this experimental feature in the framework of Regge approach
if instead of the assumption on
the Gaussian form of the cut vertex of one pomeron coupling to proton, (\ref{G}),
we will use one of the above non-Gaussian dependencies.
Of course it will cause considerable additional computing difficulties,
so we leave the realization of this program for future studies.

\section*{Acknowledgements}%
%The authors acknowledge Saint-Petersburg State University for the research grant 11.38.197.2014.

The work was supported by the Saint-Petersburg State University grant 11.38.197.2014.
V.K. acknowledges support of Special SPbSU Rector's Scholarship and
Dynasty Foundation Scholarship, and V.V. has benefited from the RFFI grant 15-02-02097.

%\begin {thebibliography}{99}
%
%\bibitem{paper1} Surname, I.I., Surname, I.I., 19??,
%            Title of the reference,
%            \emph{Journal}. Vol.\;{\bf ?}, pp.\;??--??.
%\bibitem{paper2} .....
%\end{thebibliography}

\end {document}